\documentclass[twocolumn,prl,showpacs,preprintnumbers,amsmath,amssymb,longbibliography]{revtex4-1}
\usepackage{float}
\usepackage{graphicx}
\usepackage{epsfig}
\usepackage{epstopdf}
\usepackage{dcolumn}
\usepackage{bm}
\usepackage{bm,color}
\usepackage{float}

\begin{document}
\title{Constitutive model for time-dependent flows of shear-thickening suspensions}
\author{J. J. J. Gillissen$^1$, C. Ness$^2$, J. D. Peterson$^3$, H. J. Wilson$^1$ and M. E. Cates$^3$} 
\affiliation{$^1$ Department of Mathematics, University College London, Gower Street, London WC1E 6BT, United Kingdom\\
$^2$ Department of Chemical Engineering and Biotechnology, University of Cambridge,  Cambridge CB3 0AS, United Kingdom\\
$^3$DAMTP, Centre for Mathematical Sciences, University of Cambridge, Cambridge CB3 0WA, United Kingdom}

\date{\today}
\begin{abstract}
We develop a tensorial constitutive model for dense, 
shear-thickening particle suspensions subjected to time-dependent flow. 
Our model combines a recently proposed evolution equation for the suspension microstructure in rate-independent materials with ideas developed previously to explain the steady flow of shear-thickening ones, whereby friction proliferates among compressive contacts at large particle stresses. We apply our model to shear reversal, and find good qualitative agreement with particle-level, discrete-element simulations whose results we also present.
\end{abstract}
\maketitle
Dense suspensions of solid particles occur ubiquitously in nature and industry 
\cite{guazzelli2018rheology}.
Predicting their flow behavior is essential both for understanding natural phenomena, such as mudslides and silting of waterways, and for the design of industrial products and processes ranging from paints and pharmaceuticals to chocolate 
\cite{blanco2019conching}.
At high solid concentrations, the rheology of particle suspensions differs considerably from that of a conventional Newtonian fluid. One non-Newtonian effect in many dense suspensions is a dramatic, often discontinuous, increase in viscosity with shear rate, known as shear thickening \cite{bi2011jamming,peters2016direct}.

Shear thickening is believed to
originate in a crossover from lubricated to frictional interparticle contacts 
\cite{boyer2011unifying,
pan2015s,guy2015towards,
royer2016rheological,
clavaud2017revealing,
hsiao2017rheological,
hsu2018roughness},
governed by a competition between a soft
repulsive interparticle force $F$ (of range $\epsilon \ll a$, with $a$ the hard-core particle radius), 
and the particle pressure $\Pi = -\mathrm{Tr}\boldsymbol{\Sigma}/3$ (with
$\boldsymbol\Sigma$ the macroscopic particle stress tensor).
At modest $\Pi$, the force $F(h)$ 
maintains finite separations $h$ and lubrication films
are unbroken \cite{comtet2017pairwise}.
However $\Pi$ rises with the flow rate, and
when it exceeds $\Pi^*\sim F^*/a^2$, with $F^* = \sup[F(h)]$, 
particles are pushed into frictional contact $(h\to 0)$ and lubrication films break. Frictional contacts constrain the particle dynamics, 
resulting in a rapid increase in the suspension viscosity. This can cause continuous or discontinuous shear thickening even though the underlying contact statistics always evolve smoothly with stress \cite{wyart2014discontinuous}.
This scenario has been confirmed by particle simulations, 
using the so-called `critical load model',
wherein particles experience Coulomb friction
only when their normal contact force exceeds a critical value \cite{seto2013discontinuous,mari2014shear}.

Shear thickening has been studied mainly for 
steady, homogeneous shear flow, whose behavior is well 
described by the Wyart-Cates theory (WC). This addresses
the shear viscosity 
$\eta(\phi,\dot{\gamma}) = \Sigma_{xy}/\dot\gamma$ as a function of particle volume fraction $\phi$ 
and shear rate $\dot{\gamma} = \partial_yv_x$ \cite{wyart2014discontinuous}. WC assume, with $\eta_s$ the solvent viscosity and $\nu$ some constant,
\begin{equation}
\eta=\eta_s \nu(\phi^J-\phi)^{-2}, 
\label{eq400}
\end{equation}
which diverges as $\phi\to\phi^{J}$ from below, with $\eta$ infinite beyond. Crucially, the critical value $\phi^J$ is stress-dependent, obeying~\cite{wyart2014discontinuous,guy2015towards,hermes2016unsteady} \begin{equation}
\phi^J(f)=\phi^J_1(1-f)+\phi^J_2 f\quad,\quad f(\Pi)= \exp\left(-{\Pi^*}/{\Pi}\right).
\label{eq401}
\end{equation}
Here $f(\Pi)$ is the fraction of contacts that are constrained by friction to roll, rather than slide. 

The jamming point $\phi^J$ thus evolves smoothly from a larger value $\phi^J_1$ at $\Pi \ll \Pi^*$,
to a smaller value $\phi^J_2$ for $\Pi\gg \Pi^*$. 
These limits are where frictionless and fully frictional packings become rigid. 
In interpreting (\ref{eq400},\ref{eq401}) microscopically, WC effectively assumed that the steady-state microstructure depends on $\phi$ only, which therefore measures the proximity to jamming. (Below we will need to find a more general, time-dependent `jamming coordinate'.) This requires the microstructure to be $f$-independent, whereas in reality there could be a slightly different steady-state microstructure for each $f$-value and hence for each strain rate \cite{boyer2011unifying,hermes2016unsteady}.  

The WC theory accounts for 
experimental and numerical data 
for shear thickening in steady shear flow 
\cite{guy2015towards,hermes2016unsteady, singh2018constitutive,guy2019testing}, but makes no predictions for nonstationary flows, such as the sudden reversal of steady shear. The latter gives direct access to the statistics of direct interparticle and lubrication forces; on reversal, direct repulsions can drop straight to zero (in the $\epsilon\ll a$ limit), whereas lubrication forces reverse sign at fixed magnitude \cite{gadala1980shear,lin2015hydrodynamic}. 

Extending the WC theory to nonstationary and/or non-shear flows is clearly an important task, requiring development of a tensorial constitutive equation that relates the material's state of stress to its preceding flow history. Building a new constitutive model is usually done first by assuming time-dependent but spatially homogeneous flows; spatiotemporal dynamics can later be addressed via additional terms involving spatial gradients. We take only the first step here, noting that in other soft matter systems the second step has followed only years later, see, {\em e.g.}, \cite{fielding2007complex}.

Recently, two of us (Gillissen and Wilson, GW) constructed a constitutive equation for the rheology of {\em rate-independent} suspensions \cite{gillissen2018modeling,gillissen2019effect}. Rate-independence, in which all stress components are linear in $\dot\gamma$,  arises when the frictional contact statistics are independent of flow rate: $f\neq f(\Pi)$. In this Letter we build on that work to obtain a constitutive model for shear thickening materials, exploiting the simplification already made by WC, that microstructural evolution is friction-independent. On the other hand, we allow the instantaneous relation connecting the stress tensor to the microstructure and flow rate to depend strongly on friction. Shear thickening is then captured by judiciously combining GW and WC precepts, as we describe next. 

{\em Rate-independent theory:} An evolution equation for the second rank microstructure tensor $\langle\boldsymbol{nn}\rangle$ was derived in \cite{gillissen2018modeling} from the advection equation for the distribution function $\Psi(\boldsymbol{n})$ of contact vectors $\boldsymbol{n}$ between neighbors. The unit vector $\boldsymbol{n}$ does not distinguish lubrication from direct forces; instead $\Psi(\boldsymbol{n})$ counts all particle contacts within some coarse-graining shell that is thin compared to the particle radius $a$ and thick compared to the range $\epsilon$ of the direct interparticle force $F(h)$.
The GW equation reads \cite{gillissen2018modeling}: 
\begin{multline}
\partial_t\langle \boldsymbol{nn}\rangle =
{\boldsymbol{L}}\cdot\langle \boldsymbol{nn}\rangle+\langle \boldsymbol{nn}\rangle\cdot{\boldsymbol{L}}^T-
2{\boldsymbol{L}}:\langle \boldsymbol{nnnn}\rangle\\
-\beta\left[
{\boldsymbol{E}}_e: \langle \boldsymbol{nnnn}\rangle+
\frac{
\color{black}
\phi
\color{black}
}{15}
 \left(2{\boldsymbol{E}}_c+\mathrm{Tr}({\boldsymbol{E}}_c)
 \boldsymbol{\delta}\right)
 \right].
  \label{eq2}
\end{multline}
Here $L_{ij}=\partial_j v_i$ is the velocity gradient and $v_i$ the velocity. 
The terms in $\boldsymbol{L}$ describe advection of $\boldsymbol{n}$, 
while the $\beta$-term accounts for creation and destruction 
of particle pairs. The compressive rate of strain $\boldsymbol{E}_c$  advects, into the coarse-graining shell, an isotropic exterior distribution of non-contacting particles, importing preferentially those along the compression axis or axes. In contrast the extensional rate of strain $\boldsymbol{E}_e$ advects the anisotropically distributed existing contacts out of the coarse-graining shell, exporting preferentially those along the extension axis or axes. 

Since in relatively dense systems $\Psi(\boldsymbol{n})$ is relatively close to isotropy \cite{blanc2013microstructure},
we follow GW and express the fourth moment $\langle\boldsymbol{nnnn}\rangle$ in terms of 
$\langle\boldsymbol{nn}\rangle$ via the `linear closure' \cite{hinch1976constitutive}:
\begin{multline}
\langle n_in_jn_kn_l\rangle=
-\tfrac{1}{35}\langle n_mn_m\rangle\left(\delta_{ij}\delta_{kl}+\delta_{ik}\delta_{jl}+\delta_{il}\delta_{jk}\right)\\
+\tfrac{1}{7}\Big(
\delta_{ij}\langle {n}_k{n}_l\rangle+\delta_{ik}\langle {n}_j{n}_l\rangle+\delta_{il}\langle {n}_j{n}_k\rangle\\
+\langle {n}_i{n}_j\rangle\delta_{kl}+\langle {n}_i{n}_k\rangle\delta_{jl}+\langle {n}_i{n}_l\rangle\delta_{jk}
\Big).
\label{eq3}
\end{multline}
Eqs.~(\ref{eq2},\ref{eq3}) are closed equations for microstructural evolution under arbitrary homogeneous flow. 
They merit several remarks:

{\em (i)} An unknown, order-unity coefficient in front of the $\phi$ term in \eqref{eq2} 
has been absorbed into the overall normalization of $\langle\boldsymbol{nn}\rangle$, which is allowed because, after closure, the model is linear in $\langle\boldsymbol{nn}\rangle$. This normalization is in turn absorbed into the parameters introduced in \eqref{eq1} below.

{\em (ii)} Although $\beta$ might depend on $f$, we will take $\beta$ constant so that the microstructural evolution remains independent of $\boldsymbol{\Sigma}$ during shear thickening.

{\em (iii)} Crucially, the model is nonlinear in $\boldsymbol{E} = \boldsymbol{E}_c+\boldsymbol{E}_e$, but separately linear in 
$\boldsymbol{E}_c$ and $\boldsymbol{E}_e$; these uniquely decompose $\boldsymbol{E} = (\boldsymbol{L} + \boldsymbol{L}^T)/2$ into its positive and negative eigen-parts. This {\em piecewise linearity} places the model outside a linear class that was found inadequate for flow reversal modeling \cite{chacko2018shear}, 
while avoiding the proliferating parameters of general nonlinearity. Frame invariance remains encoded in the advective terms of Eq.~\eqref{eq2}~\cite{hinch1976constitutive}.

{\em (iv)} Eqs.~(\ref{eq2},\ref{eq3}) 
predict unphysical oscillations for $\beta\leq 3$ 
in simple shear flow \cite{gillissen2018modeling,gillissen2019effect},
so we restrict to $\beta>3$.

To complete their rate-independent model, GW adopted an instantaneous relation between particle stress, microstructure and strain rate \cite{gillissen2019effect}:
\color{black}
\begin{equation}
\boldsymbol{\Sigma}=
\eta_s
\left[
{\alpha \boldsymbol{E}}
+{\chi\boldsymbol{E}_c}
\right]
:\langle\boldsymbol{nnnn}\rangle.
\label{eq1}
\end{equation}
Here the $\alpha$-term represents lubrication forces, and
the $\chi$-term direct interparticle forces ($F(h)$, hard-core repulsions, and friction); all tangential contributions are omitted as subdominant 
\cite{seto2018normal}.
Importantly, on flow reversal  $\boldsymbol{E}_c$ and $\boldsymbol{E}_e$ interchange, so that \eqref{eq1} captures the discontinuous drop in particle stress as direct contacts, oriented mainly along the previously compressive axis, suddenly open. In contrast, as required by Stokesian reversibility, the lubrication part changes sign at fixed magnitude on reversal \cite{gadala1980shear}.

GW showed that Eqs.~(\ref{eq2}-\ref{eq1}) 
predict qualitatively correct results for stress and microstructure in suspensions of rate-independent rheology, 
for both steady and reversing flows \cite{gillissen2018modeling,gillissen2019effect}. 
The model also correctly predicts the destabilising effect of spheres on Taylor vortices \cite{gillissen2019c}. 

{\em Constitutive model for shear thickening:} Our task is to marry these results for rate-independent materials to the physics of shear thickening as described by WC theory \cite{wyart2014discontinuous}. To achieve this we should allow the stress parameter $\chi$ in (\ref{eq1}) to depend on the fraction $f$ of direct contacts that are frictional, which evolves from mostly frictionless ($f\simeq 0$) to mostly frictional ($f\simeq 1$) as $\Pi$ grows beyond $\Pi^*$. However it is no longer possible to replace the dependence of viscosity on microstructure with a dependence on $\phi-\phi^J(f)$ as done in (\ref{eq400},\ref{eq401}). This is because the microstructure, unlike  $\phi$, evolves in time. 

We therefore need to identify within the model a `jamming coordinate' $\xi$ that estimates, for a given microstructure and flow, the system's distance from a jamming point $\xi^J(\Pi)$. One candidate for $\xi$ is $\mathrm{Tr}\langle\boldsymbol{nn}\rangle$ which (up to a prefactor, see remark {\em (i)} above) counts all contacts within the coarse-graining shell. But only a subset of these (those within the range $\epsilon$ of direct interactions) are candidates for becoming frictional; and the same coarse-grained microstructure could be near to, or far from, jamming depending on the flow geometry \cite{cates1998jamming}.

Since these direct contacts are mainly orientated along the compression axis/axes we adopt as our jamming coordinate the contraction of the microstructure onto $\boldsymbol{E}_c$:
\begin{equation}
\xi\equiv-\frac{\langle\boldsymbol{nn}\rangle:\boldsymbol{E}_c}{\sqrt{\boldsymbol{E}_c:\boldsymbol{E}_c}}.
\label{eq40}
\end{equation}
We show below that, in particle simulations, $\xi$ evolves similarly to a coordination number $Z$ that counts direct ($h<\epsilon$) contacts only. This $Z$ might be an equally good choice for the jamming coordinate \cite{wyart2014discontinuous}, but it is not calculable within our coarse-grained constitutive model.

The jamming point for $\xi$, denoted $\xi^J(f)$, must decrease from a larger value $\xi^J_1$ to a smaller value $\xi^J_2$ as friction switches on. Following \eqref{eq401} we write:
\begin{equation}
\xi^J(f) =(1-f)\xi^J_1+f\xi^J_2\quad,\quad f(\Pi) = \exp(-\Pi^*/\Pi).
\label{eq8}
\end{equation}
To find the extremal jamming points  $\xi^J_{1,2}$, 
we denote by $\xi_{\infty}\left(\phi,\beta,\boldsymbol{L}\right)$
the steady state solution of  (\ref{eq2}, \ref{eq3}, \ref{eq40}) 
at given velocity gradient $\boldsymbol{L}$. 
Supposing the critical volume fractions
$\phi^J_{1,2}$ to be known, as they are for simple shear flows,  we can then identify $\xi^J_{1,2}=\xi_{\infty}\left(\phi^J_{1,2},\beta,\boldsymbol{L}\right)$. 

\begin{figure}
\includegraphics[width=\linewidth] {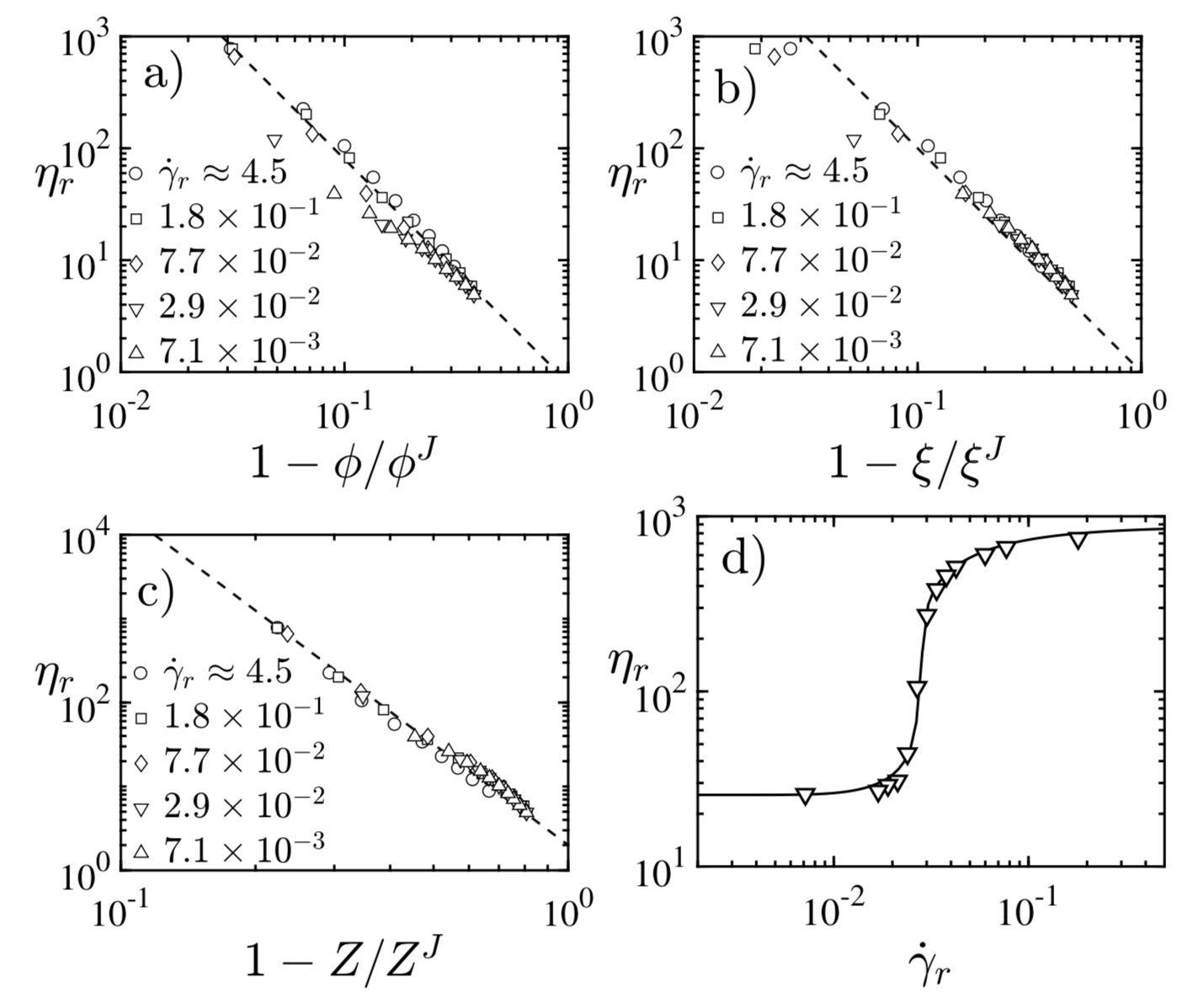}
\caption{
(a) DEM results for $\eta_r$ 
as a function of $1-\phi/\phi^J$ [where $\phi^J$ obeys (\ref{eq401})]
for various $\dot{\gamma}_r$.
Dashed line: slope $-2$.
(b) The same data plotted against $1-\xi/\xi^J$.
Dashed line:  slope $-2$. 
(c) The same data plotted against $1-Z/Z^J$. 
Dashed line: slope $-4$. 
(d) Steady state flow curve $\eta_r(\dot\gamma_r)$ for $\phi = 0.56$. 
Solid line: fitted model.
}
\label{fig6}
\end{figure}

We finally assign the dependence of $\chi$ in (\ref{eq1}) on $\xi$:
\begin{equation}
\chi={\chi_0}{\left(1-{\xi}/{\xi^J}\right)^{-2}},
\label{eq31}
\end{equation}
where the exponent $-2$ is justified by our particle simulations; see Fig.~\ref{fig6}b below. This is the same exponent as in (\ref{eq400}), so that $\xi-\xi^J$ emerges as a direct dynamical counterpart of $\phi-\phi^J$ in WC theory. (Using $Z-Z^J$ would entail a different exponent in \eqref{eq31}; see Fig.~\ref{fig6}c.)

Eqs.~(\ref{eq2}-\ref{eq31}) define our constitutive model.
They contain the parameters $\phi$, $\phi^J_{1,2}$, $\Pi^*$, $\chi_0$, $\alpha$ and $\beta$ of which the first four are already present in the WC theory -- with our $\chi_0$ replacing $\nu$ in \eqref{eq401}. Thus our model extends the WC predictions from steady shear to arbitrary, unsteady but homogeneous flow, 
at the cost of just two new parameters. Of these, $\alpha$ governs the lubrication stress, subdominant near the frictional jamming point and omitted by WC. Time dependence is controlled by $\beta$, which gives a strain scale for structural evolution via \eqref{eq2}. Although $\alpha$ and $\beta$ depend on $\phi$, they should diverge only on approach to $\phi_1$, so are near-constant in the neighbourhood of $\phi_2$. For simplicity we fit them below to simulation data at  $\phi = 0.56$.  

{\em DEM simulations:} We now test our predictions against simulations using
the discrete element method (DEM)  \cite{plimpton1995fast,cheal2018rheology}. 
We use equimolar bidisperse spheres with
density $\rho$, radii $a$ and $1.4a$ and volume 
fraction $\phi$
in a periodic box at imposed shear rate $\dot{\gamma}$.
These particles obey Newton's laws with short-range, pairwise (centre-to-centre unit vector $\boldsymbol{n}$) interactions.  Lubrication forces 
\cite{ball1997simulation} act at separations below $0.05a$; 
direct forces obey
${\bm F} = k_n\delta{\bm n} - k_t{\bm t}$,
for overlap $\delta$, stiffnesses $k_n$ and $k_t$, and tangential displacement ${\bm t}$.
The tangential force is restricted by a friction coefficient $\mu$ 
so that $|k_t {\bm t}| \leq \mu k_n\delta$.
The suspension stress is found 
by summing all hydrodynamic and contact stresslets.
Choosing $\rho\dot{\gamma}a^2 = 10^{-3}\eta_s$ and 
$\dot{\gamma} = 10^{-5}\sqrt{k_n/(2\rho a)}$, we approach
inertialess, hard sphere conditions,
and match experiments on rate-independent rheology \cite{boyer2011unifying}.
Shear thickening is then added using the `critical load model' 
\cite{seto2013discontinuous}: contacts with ${\bm F}\cdot{\bm n} > F^*$
have $\mu=1$, others have $\mu=0$.
The frictional crossover is then governed 
by a reduced shear rate
$\dot{\gamma}_r =\dot{\gamma}\eta_s/\Pi^*\sim\eta_s a^2/F^*$. Results are averaged over 40 simulations, each containing 1500 particles. This system size is large enough to give detailed microstructural statistics but small enough to maintain uniformity of the particle density 
\cite{chacko2018dynamic}.
To calculate $\xi$ we take a coarse-graining shell thickness set by the lubrication range $(0.05a)$, whereas $Z$ is found by counting overlapping particles only.

{\em Steady-state results:} According to our model, 
for a given material, 
the reduced viscosity $\eta_r = \Sigma_{xy}/(\dot\gamma\eta_s)$ is a function of $\dot\gamma_r$ (defined above) and $\phi$ only. In steady state, where \eqref{eq400} works well 
\cite{guy2015towards, hermes2016unsteady}, $\eta_r$ should depend mainly on the distance of $\phi$ from $\phi^J(f)$, which varies with $\dot\gamma_r$ via $f(\Pi)$. We test this
using our DEM data
by plotting in Fig.~\ref{fig6}a, on log-log axes, $\eta_r$ against $1-\phi/\phi^J(f)$ for various $\dot\gamma_r$ and $0.4\le\phi\le0.64$. With $\phi^J_1 = 0.644, \phi^J_2 = 0.578$  and $\Pi^* = 0.037 F^*/a^2$, there is good data collapse, with slope of $-2$, confirming the exponent chosen in \eqref{eq400} above \cite{wyart2014discontinuous}. 
In Fig.~\ref{fig6}b the same data are plotted against $1-\xi/\xi^J(f)$, with $\xi,\xi^J(f)$ obeying (\ref{eq40},\ref{eq8}). For these purposes, jamming points $\xi^J_{1,2} = 2.615, 2.069$ were found by plotting $\xi$ against $Z$ and reading off values for frictionless and frictional jamming ($Z = 6,4$). 
 The collapse quality is comparable to Fig.~\ref{fig6}a, with the same exponent, confirming \eqref{eq31}. A similar plot using $Z$ as the ordinate instead gives an exponent $-4$, see Fig.~\ref{fig6}c (for more on the $Z$--$\phi$ relationship see 
 \cite{radhakrishnan2019force}).

Fig.~\ref{fig6}d compares our model with DEM results for a steady-state flow curve $\eta_r(\dot\gamma_r)$, 
at volume fraction $\phi=0.56$, within the regime of continuous shear thickening. Parameters $\Pi^*$ and $\xi^J_{1,2} = 0.88, 0.78$ were found as previously described, assuming $\phi_{1,2}=0.65,0.57$. 
(The latter, found via $\xi^J_{1,2}=\xi_{\infty}\left(\phi^J_{1,2},\beta,\boldsymbol{L}\right)$, absorb a  normalization ---see remark {\em (i)} above--- so are not directly comparable with simulation values.) The curve is well fit with  $\alpha=120$,  and $\chi_0=2.4$. In choosing the above parameters, we hold $\beta = 50$; this is fitted to microstructural evolution data following shear reversal, described next.

\begin{figure}
\includegraphics[width=\linewidth] {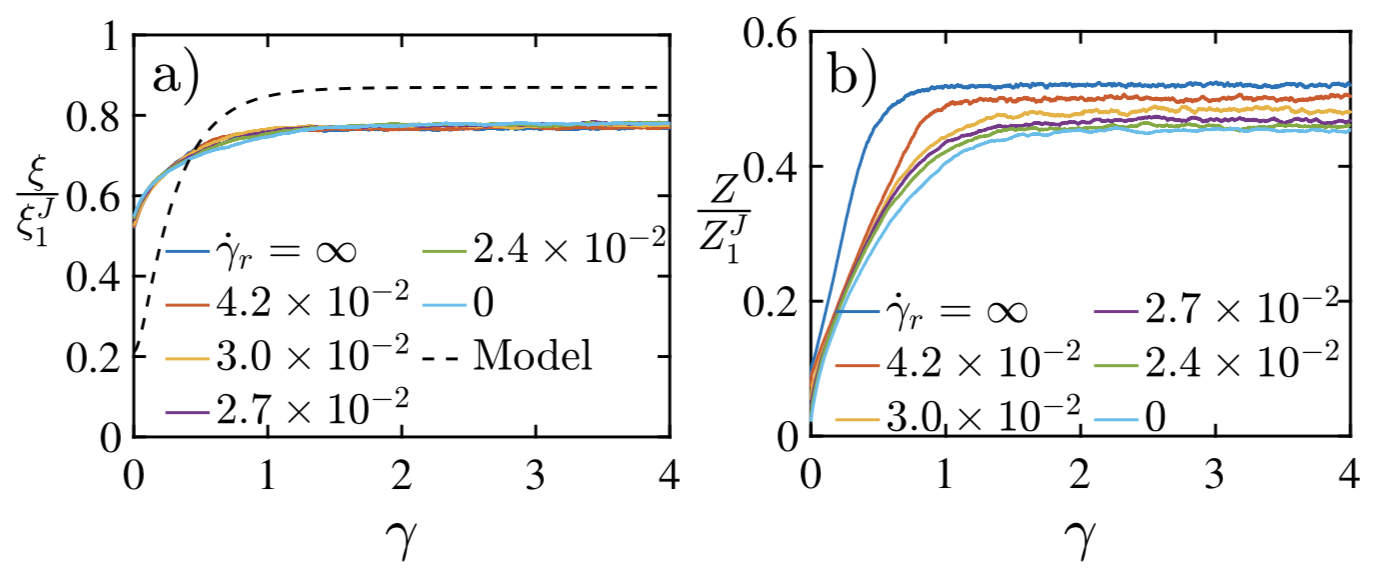}
\caption{ 
\color{black}
(a) Evolution of jamming coordinate $\xi$ via DEM after shear reversal for various reduced strain rates
$\dot\gamma_r$ at  $\phi=0.56$. Dashed curve: prediction of the model for $\beta = 50$. (b) Similar plot for the
coordination number $Z$ via DEM.
\color{black}
}
\label{fig4}
\end{figure}

\begin{figure}
\includegraphics[width=\linewidth] {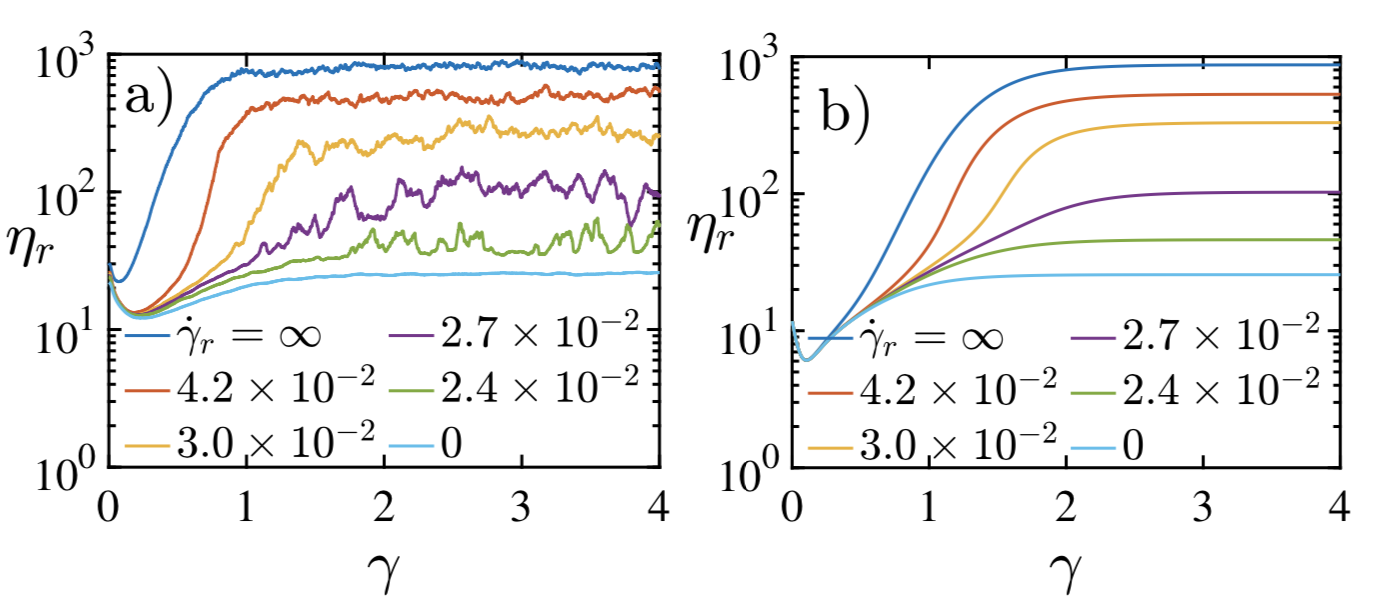}
\caption{ 
\color{black}
Reduced suspension viscosity $\eta_r$
after shear reversal, 
for various reduced shear rates $\dot{\gamma}_r$ 
at $\phi=0.56$,
computed with (a) the DEM and (b) the constitutive model.
\color{black}
}
\label{fig7}
\end{figure}

{\em Shear reversal:}
In this protocol
the suspension is sheared with negative $\dot\gamma$ until steady state is reached; at $t=0$ the flow is reversed. 
In steady state, pre-reversal, the contact vectors $\boldsymbol{n}$ are primarily aligned with the compression axis. On reversal, the compression and extensional axes interchange. 
Extensional flow then pulls contacts apart, 
decreasing both $\xi$ and $Z$ discontinuously.
This is followed by recovery, as
contacts re-form
along the new compression axis.

Fig.~\ref{fig4}a shows, for $\phi = 0.56$,  the time evolution of  the jamming coordinate $\xi$, scaled by the frictionless jamming point $\xi^J_1$, as a function of the forward strain $\gamma=t\dot{\gamma}$ after reversal, for various reduced shear rates $\dot\gamma_r$. 
(The limiting cases of $\dot\gamma_r = 0,\infty$ correspond to frictionless and frictional rate-independent materials.) 
We set $\beta = 50$ to match the observed strain scale for recovery, giving the model curve shown by the dashed line. 
Our model predicts a single curve for $\xi(\gamma)$ because it assumes 
that the microstructural evolution is not itself friction-dependent. 
This is supported by the DEM data. 
Fig.~\ref{fig4}b shows $Z$ in place of $\xi$, 
giving similar behavior but indicating weak rate-dependence of 
the direct particle contacts, not resolved by our model. 
Our rate-independent coarse-grained microstructure allows us to fit $\beta$ 
without knowledge of the stress. 
Time-dependent stress measurements can then test our model with its parameters fixed by {\em separate data} drawn from the steady-state stress and microstructural reversal results (Figs.~\ref{fig6}a, d, \ref{fig4}a).  

In Fig.~\ref{fig7} we show such a test, using DEM data for shear viscosity after reversal. (Note that laboratory measurements broadly agree with DEM \cite{lin2015hydrodynamic,blanc2018universal}.)
Our model predicts that upon reversal the viscosity $\eta_r$ drops discontinuously,
and then recovers gradually to the steady-state value. It captures remarkably well the DEM data, even though the actual DEM dynamics at small strain scales is more complex: first the direct contact stress drops to almost zero over a tiny strain interval, followed by a surge in lubrication stress at strains $\gamma \le 10^{-2}$ caused by rapid separation of particle pairs \cite{ness2016two}. Without resolving this fast regime our model captures well the subsequent evolution of both quantities: a drop in lubrication stress over strains of order $0.2$ is compensated only later by the recovery of direct contact stress, explaining the initial dip in the curves.

\begin{figure}
\includegraphics[width=\linewidth] {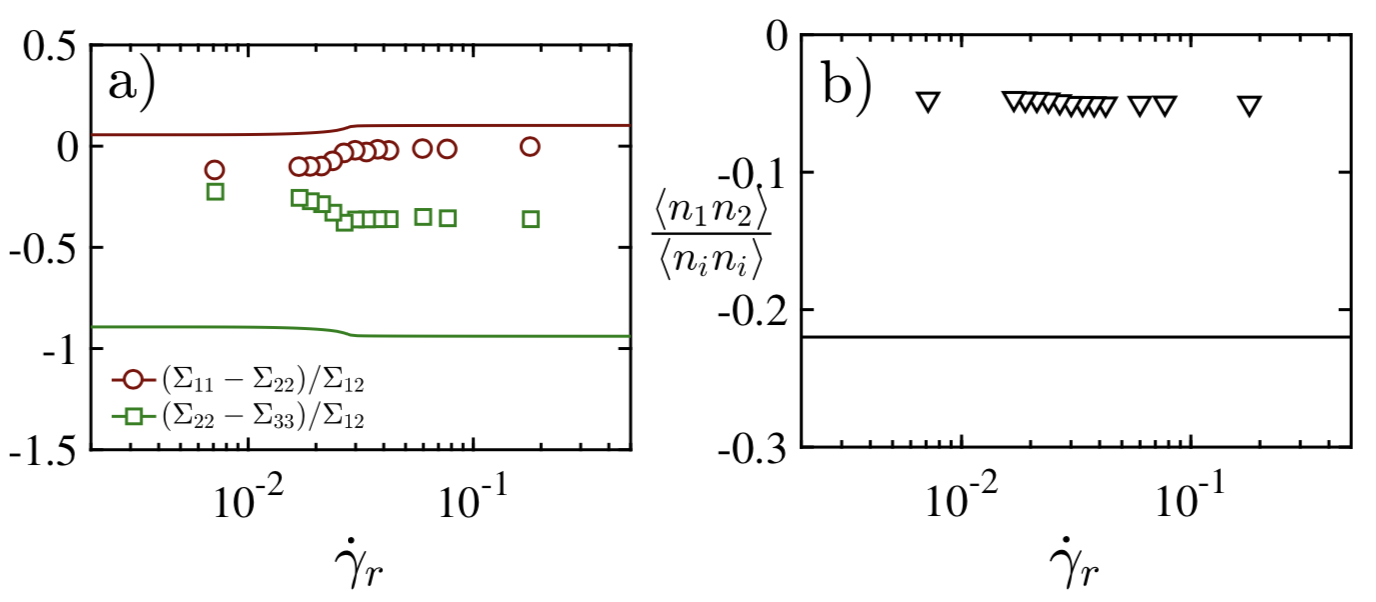}
\caption{Steady-state normal stress ratios (a) and 
microstructure anisotropy (b), versus reduced shear rate. 
Comparison between model (lines) and DEM (symbols). }
\label{fig2}
\end{figure}

{\em Normal stresses and anisotropy:} Alongside its ability to treat dynamics, our model (unlike WC theory) predicts the full stress tensor. Fig.~\ref{fig2}a shows results for the normal stress ratios in steady shear as functions of shear rate. The second normal stress ratio, $(\Sigma_{yy}-\Sigma_{zz})/\Sigma_{xy}$, is negative in both cases, as in experiments \cite{denn2014rheology,cwalina2014material}. The DEM results show an increase on thickening; our model over-predicts the value, and under-predicts this increase. This reflects a general over-prediction of
microstructural anisotropy in the model, causing too big a discontinuous drop in $\xi$ on reversal (Fig.~\ref{fig4}a), and too negative a steady-state value of $\langle n_1n_2\rangle/\langle n_in_i\rangle$ (Fig.~\ref{fig2}b). A possible cause is that, in modelling birth-and-death terms, \eqref{eq2} does not account for the effects of steric hindrance in limiting anisotropy.  Note also that the first normal stress ratio $(\Sigma_{xx}-\Sigma_{yy})/\Sigma_{xy}$ is weakly positive in the model with a small change on thickening, but negative (and almost zero when thickened) in DEM. However, this small ratio is notoriously elusive for both prediction and experiment; even its sign is controversial \cite{denn2014rheology}.

{\em Conclusions:} We have created a tensorial constitutive model for shear thickening suspensions in time-dependent flows.
Our model assumes rate-independent microstructural evolution \cite{gillissen2018modeling}, but introduces a time-dependent jamming coordinate $\xi$ that tracks the distance from a jamming point $\xi^J(\Pi)$, encoding the proliferation of contact friction at high particle pressure $\Pi$ \cite{wyart2014discontinuous}. Marrying these elements, and with parameters fit using separate data, the model successfully predicts the time-dependent shear stress after strain reversal -- with a discontinuous drop as direct contacts are lost, followed by a further gentle decline as lubrication contacts weaken, before both types of contacts rebuild and steady state is restored. The model opens several avenues for future work, such as an account of how friction feeds back into the microstructure, and a better account of saturating anisotropy, which should quantitatively improve its rheological predictions.

{\em Acknowledgements}: We acknowledge financial support from the Engineering and Physical Sciences Research Council of the United Kingdom Grant No. EP/N024915/1, and from the European Research Council under the Horizon 2020 Programme, ERC grant agreement number 740269. MEC is funded by the Royal Society. CN is funded by the Maudslay-Butler Research Fellowship at Pembroke College, Cambridge.
\bibliography{article}

\begin{thebibliography}{38}%
\makeatletter
\providecommand \@ifxundefined [1]{%
 \@ifx{#1\undefined}
}%
\providecommand \@ifnum [1]{%
 \ifnum #1\expandafter \@firstoftwo
 \else \expandafter \@secondoftwo
 \fi
}%
\providecommand \@ifx [1]{%
 \ifx #1\expandafter \@firstoftwo
 \else \expandafter \@secondoftwo
 \fi
}%
\providecommand \natexlab [1]{#1}%
\providecommand \enquote  [1]{``#1''}%
\providecommand \bibnamefont  [1]{#1}%
\providecommand \bibfnamefont [1]{#1}%
\providecommand \citenamefont [1]{#1}%
\providecommand \href@noop [0]{\@secondoftwo}%
\providecommand \href [0]{\begingroup \@sanitize@url \@href}%
\providecommand \@href[1]{\@@startlink{#1}\@@href}%
\providecommand \@@href[1]{\endgroup#1\@@endlink}%
\providecommand \@sanitize@url [0]{\catcode `\\12\catcode `\$12\catcode
  `\&12\catcode `\#12\catcode `\^12\catcode `\_12\catcode `\%12\relax}%
\providecommand \@@startlink[1]{}%
\providecommand \@@endlink[0]{}%
\providecommand \url  [0]{\begingroup\@sanitize@url \@url }%
\providecommand \@url [1]{\endgroup\@href {#1}{\urlprefix }}%
\providecommand \urlprefix  [0]{URL }%
\providecommand \Eprint [0]{\href }%
\providecommand \doibase [0]{http://dx.doi.org/}%
\providecommand \selectlanguage [0]{\@gobble}%
\providecommand \bibinfo  [0]{\@secondoftwo}%
\providecommand \bibfield  [0]{\@secondoftwo}%
\providecommand \translation [1]{[#1]}%
\providecommand \BibitemOpen [0]{}%
\providecommand \bibitemStop [0]{}%
\providecommand \bibitemNoStop [0]{.\EOS\space}%
\providecommand \EOS [0]{\spacefactor3000\relax}%
\providecommand \BibitemShut  [1]{\csname bibitem#1\endcsname}%
\let\auto@bib@innerbib\@empty
\bibitem [{\citenamefont {Guazzelli}\ and\ \citenamefont
  {Pouliquen}(2018)}]{guazzelli2018rheology}%
  \BibitemOpen
  \bibfield  {author} {\bibinfo {author} {\bibfnamefont {{\'E}.}~\bibnamefont
  {Guazzelli}}\ and\ \bibinfo {author} {\bibfnamefont {O.}~\bibnamefont
  {Pouliquen}},\ }\bibfield  {title} {\enquote {\bibinfo {title} {Rheology of
  dense granular suspensions},}\ }\href@noop {} {\bibfield  {journal} {\bibinfo
   {journal} {J. Fluid Mech.}\ }\textbf {\bibinfo {volume} {852}} (\bibinfo
  {year} {2018})}\BibitemShut {NoStop}%
\bibitem [{\citenamefont {Blanco}\ \emph {et~al.}(2019)\citenamefont {Blanco},
  \citenamefont {Hodgson}, \citenamefont {Hermes}, \citenamefont {Besseling},
  \citenamefont {Hunter}, \citenamefont {Chaikin}, \citenamefont {Cates},
  \citenamefont {Van~Damme},\ and\ \citenamefont {Poon}}]{blanco2019conching}%
  \BibitemOpen
  \bibfield  {author} {\bibinfo {author} {\bibfnamefont {E.}~\bibnamefont
  {Blanco}}, \bibinfo {author} {\bibfnamefont {D.~J.~M.}\ \bibnamefont
  {Hodgson}}, \bibinfo {author} {\bibfnamefont {M.}~\bibnamefont {Hermes}},
  \bibinfo {author} {\bibfnamefont {R.}~\bibnamefont {Besseling}}, \bibinfo
  {author} {\bibfnamefont {G.~L.}\ \bibnamefont {Hunter}}, \bibinfo {author}
  {\bibfnamefont {P.~M.}\ \bibnamefont {Chaikin}}, \bibinfo {author}
  {\bibfnamefont {M.~E.}\ \bibnamefont {Cates}}, \bibinfo {author}
  {\bibfnamefont {I.}~\bibnamefont {Van~Damme}}, \ and\ \bibinfo {author}
  {\bibfnamefont {W.~C.~K.}\ \bibnamefont {Poon}},\ }\bibfield  {title}
  {\enquote {\bibinfo {title} {Conching chocolate is a prototypical transition
  from frictionally jammed solid to flowable suspension with maximal solid
  content},}\ }\href@noop {} {\bibfield  {journal} {\bibinfo  {journal} {P.
  Natl. Acad. Sci.}\ }\textbf {\bibinfo {volume} {116}},\ \bibinfo {pages}
  {10303--10308} (\bibinfo {year} {2019})}\BibitemShut {NoStop}%
\bibitem [{\citenamefont {Bi}\ \emph {et~al.}(2011)\citenamefont {Bi},
  \citenamefont {Zhang}, \citenamefont {Chakraborty},\ and\ \citenamefont
  {Behringer}}]{bi2011jamming}%
  \BibitemOpen
  \bibfield  {author} {\bibinfo {author} {\bibfnamefont {D.}~\bibnamefont
  {Bi}}, \bibinfo {author} {\bibfnamefont {J.}~\bibnamefont {Zhang}}, \bibinfo
  {author} {\bibfnamefont {B.}~\bibnamefont {Chakraborty}}, \ and\ \bibinfo
  {author} {\bibfnamefont {R.~P.}\ \bibnamefont {Behringer}},\ }\bibfield
  {title} {\enquote {\bibinfo {title} {Jamming by shear},}\ }\href@noop {}
  {\bibfield  {journal} {\bibinfo  {journal} {Nature}\ }\textbf {\bibinfo
  {volume} {480}},\ \bibinfo {pages} {355} (\bibinfo {year}
  {2011})}\BibitemShut {NoStop}%
\bibitem [{\citenamefont {Peters}\ \emph {et~al.}(2016)\citenamefont {Peters},
  \citenamefont {Majumdar},\ and\ \citenamefont {Jaeger}}]{peters2016direct}%
  \BibitemOpen
  \bibfield  {author} {\bibinfo {author} {\bibfnamefont {I.~R.}\ \bibnamefont
  {Peters}}, \bibinfo {author} {\bibfnamefont {S.}~\bibnamefont {Majumdar}}, \
  and\ \bibinfo {author} {\bibfnamefont {H.~M.}\ \bibnamefont {Jaeger}},\
  }\bibfield  {title} {\enquote {\bibinfo {title} {Direct observation of
  dynamic shear jamming in dense suspensions},}\ }\href@noop {} {\bibfield
  {journal} {\bibinfo  {journal} {Nature}\ }\textbf {\bibinfo {volume} {532}},\
  \bibinfo {pages} {214} (\bibinfo {year} {2016})}\BibitemShut {NoStop}%
\bibitem [{\citenamefont {Boyer}\ \emph {et~al.}(2011)\citenamefont {Boyer},
  \citenamefont {Guazzelli},\ and\ \citenamefont
  {Pouliquen}}]{boyer2011unifying}%
  \BibitemOpen
  \bibfield  {author} {\bibinfo {author} {\bibfnamefont {F.}~\bibnamefont
  {Boyer}}, \bibinfo {author} {\bibfnamefont {{\'E}.}~\bibnamefont
  {Guazzelli}}, \ and\ \bibinfo {author} {\bibfnamefont {O.}~\bibnamefont
  {Pouliquen}},\ }\bibfield  {title} {\enquote {\bibinfo {title} {Unifying
  suspension and granular rheology},}\ }\href@noop {} {\bibfield  {journal}
  {\bibinfo  {journal} {Phys. Rev. Lett.}\ }\textbf {\bibinfo {volume} {107}},\
  \bibinfo {pages} {188301} (\bibinfo {year} {2011})}\BibitemShut {NoStop}%
\bibitem [{\citenamefont {Pan}\ \emph {et~al.}(2015)\citenamefont {Pan},
  \citenamefont {de~Cagny}, \citenamefont {Weber},\ and\ \citenamefont
  {Bonn}}]{pan2015s}%
  \BibitemOpen
  \bibfield  {author} {\bibinfo {author} {\bibfnamefont {Z.}~\bibnamefont
  {Pan}}, \bibinfo {author} {\bibfnamefont {H.}~\bibnamefont {de~Cagny}},
  \bibinfo {author} {\bibfnamefont {B.}~\bibnamefont {Weber}}, \ and\ \bibinfo
  {author} {\bibfnamefont {D.}~\bibnamefont {Bonn}},\ }\bibfield  {title}
  {\enquote {\bibinfo {title} {S-shaped flow curves of shear thickening
  suspensions: Direct observation of frictional rheology},}\ }\href@noop {}
  {\bibfield  {journal} {\bibinfo  {journal} {Phys. Rev. E}\ }\textbf {\bibinfo
  {volume} {92}},\ \bibinfo {pages} {032202} (\bibinfo {year}
  {2015})}\BibitemShut {NoStop}%
\bibitem [{\citenamefont {Guy}\ \emph {et~al.}(2015)\citenamefont {Guy},
  \citenamefont {Hermes},\ and\ \citenamefont {Poon}}]{guy2015towards}%
  \BibitemOpen
  \bibfield  {author} {\bibinfo {author} {\bibfnamefont {B.~M.}\ \bibnamefont
  {Guy}}, \bibinfo {author} {\bibfnamefont {M.}~\bibnamefont {Hermes}}, \ and\
  \bibinfo {author} {\bibfnamefont {W.~C.~K.}\ \bibnamefont {Poon}},\
  }\bibfield  {title} {\enquote {\bibinfo {title} {Towards a unified
  description of the rheology of hard-particle suspensions},}\ }\href@noop {}
  {\bibfield  {journal} {\bibinfo  {journal} {Phys. Rev. Lett.}\ }\textbf
  {\bibinfo {volume} {115}},\ \bibinfo {pages} {088304} (\bibinfo {year}
  {2015})}\BibitemShut {NoStop}%
\bibitem [{\citenamefont {Royer}\ \emph {et~al.}(2016)\citenamefont {Royer},
  \citenamefont {Blair},\ and\ \citenamefont {Hudson}}]{royer2016rheological}%
  \BibitemOpen
  \bibfield  {author} {\bibinfo {author} {\bibfnamefont {J.~R.}\ \bibnamefont
  {Royer}}, \bibinfo {author} {\bibfnamefont {D.~L.}\ \bibnamefont {Blair}}, \
  and\ \bibinfo {author} {\bibfnamefont {S.~D.}\ \bibnamefont {Hudson}},\
  }\bibfield  {title} {\enquote {\bibinfo {title} {Rheological signature of
  frictional interactions in shear thickening suspensions},}\ }\href@noop {}
  {\bibfield  {journal} {\bibinfo  {journal} {Phys. Rev. Lett.}\ }\textbf
  {\bibinfo {volume} {116}},\ \bibinfo {pages} {188301} (\bibinfo {year}
  {2016})}\BibitemShut {NoStop}%
\bibitem [{\citenamefont {Clavaud}\ \emph {et~al.}(2017)\citenamefont
  {Clavaud}, \citenamefont {B{\'e}rut}, \citenamefont {Metzger},\ and\
  \citenamefont {Forterre}}]{clavaud2017revealing}%
  \BibitemOpen
  \bibfield  {author} {\bibinfo {author} {\bibfnamefont {C.}~\bibnamefont
  {Clavaud}}, \bibinfo {author} {\bibfnamefont {A.}~\bibnamefont {B{\'e}rut}},
  \bibinfo {author} {\bibfnamefont {B.}~\bibnamefont {Metzger}}, \ and\
  \bibinfo {author} {\bibfnamefont {Y.}~\bibnamefont {Forterre}},\ }\bibfield
  {title} {\enquote {\bibinfo {title} {Revealing the frictional transition in
  shear-thickening suspensions},}\ }\href@noop {} {\bibfield  {journal}
  {\bibinfo  {journal} {P. Natl. Acad. Sci.}\ }\textbf {\bibinfo {volume}
  {114}},\ \bibinfo {pages} {5147--5152} (\bibinfo {year} {2017})}\BibitemShut
  {NoStop}%
\bibitem [{\citenamefont {Hsiao}\ \emph {et~al.}(2017)\citenamefont {Hsiao},
  \citenamefont {Jamali}, \citenamefont {Glynos}, \citenamefont {Green},
  \citenamefont {Larson},\ and\ \citenamefont
  {Solomon}}]{hsiao2017rheological}%
  \BibitemOpen
  \bibfield  {author} {\bibinfo {author} {\bibfnamefont {L.~C.}\ \bibnamefont
  {Hsiao}}, \bibinfo {author} {\bibfnamefont {S.}~\bibnamefont {Jamali}},
  \bibinfo {author} {\bibfnamefont {E.}~\bibnamefont {Glynos}}, \bibinfo
  {author} {\bibfnamefont {P.~F.}\ \bibnamefont {Green}}, \bibinfo {author}
  {\bibfnamefont {R.~G.}\ \bibnamefont {Larson}}, \ and\ \bibinfo {author}
  {\bibfnamefont {M.~J.}\ \bibnamefont {Solomon}},\ }\bibfield  {title}
  {\enquote {\bibinfo {title} {Rheological state diagrams for rough colloids in
  shear flow},}\ }\href@noop {} {\bibfield  {journal} {\bibinfo  {journal}
  {Phys. Rev. Lett.}\ }\textbf {\bibinfo {volume} {119}},\ \bibinfo {pages}
  {158001} (\bibinfo {year} {2017})}\BibitemShut {NoStop}%
\bibitem [{\citenamefont {Hsu}\ \emph {et~al.}(2018)\citenamefont {Hsu},
  \citenamefont {Ramakrishna}, \citenamefont {Zanini}, \citenamefont
  {Spencer},\ and\ \citenamefont {Isa}}]{hsu2018roughness}%
  \BibitemOpen
  \bibfield  {author} {\bibinfo {author} {\bibfnamefont {C.~P.}\ \bibnamefont
  {Hsu}}, \bibinfo {author} {\bibfnamefont {S.~N.}\ \bibnamefont
  {Ramakrishna}}, \bibinfo {author} {\bibfnamefont {M.}~\bibnamefont {Zanini}},
  \bibinfo {author} {\bibfnamefont {N.~D.}\ \bibnamefont {Spencer}}, \ and\
  \bibinfo {author} {\bibfnamefont {L.}~\bibnamefont {Isa}},\ }\bibfield
  {title} {\enquote {\bibinfo {title} {Roughness-dependent tribology effects on
  discontinuous shear thickening},}\ }\href@noop {} {\bibfield  {journal}
  {\bibinfo  {journal} {P. Natl. Acad. Sci.}\ }\textbf {\bibinfo {volume}
  {115}},\ \bibinfo {pages} {5117--5122} (\bibinfo {year} {2018})}\BibitemShut
  {NoStop}%
\bibitem [{\citenamefont {Comtet}\ \emph {et~al.}(2017)\citenamefont {Comtet},
  \citenamefont {Chatt{\'e}}, \citenamefont {Nigu{\`e}s}, \citenamefont
  {Bocquet}, \citenamefont {Siria},\ and\ \citenamefont
  {Colin}}]{comtet2017pairwise}%
  \BibitemOpen
  \bibfield  {author} {\bibinfo {author} {\bibfnamefont {J.}~\bibnamefont
  {Comtet}}, \bibinfo {author} {\bibfnamefont {G.}~\bibnamefont {Chatt{\'e}}},
  \bibinfo {author} {\bibfnamefont {A.}~\bibnamefont {Nigu{\`e}s}}, \bibinfo
  {author} {\bibfnamefont {L.}~\bibnamefont {Bocquet}}, \bibinfo {author}
  {\bibfnamefont {A.}~\bibnamefont {Siria}}, \ and\ \bibinfo {author}
  {\bibfnamefont {A.}~\bibnamefont {Colin}},\ }\bibfield  {title} {\enquote
  {\bibinfo {title} {Pairwise frictional profile between particles determines
  discontinuous shear thickening transition in non-colloidal suspensions},}\
  }\href@noop {} {\bibfield  {journal} {\bibinfo  {journal} {Nature Commun.}\
  }\textbf {\bibinfo {volume} {8}},\ \bibinfo {pages} {15633} (\bibinfo {year}
  {2017})}\BibitemShut {NoStop}%
\bibitem [{\citenamefont {Wyart}\ and\ \citenamefont
  {Cates}(2014)}]{wyart2014discontinuous}%
  \BibitemOpen
  \bibfield  {author} {\bibinfo {author} {\bibfnamefont {M.}~\bibnamefont
  {Wyart}}\ and\ \bibinfo {author} {\bibfnamefont {M.~E.}\ \bibnamefont
  {Cates}},\ }\bibfield  {title} {\enquote {\bibinfo {title} {Discontinuous
  shear thickening without inertia in dense non-{B}rownian suspensions},}\
  }\href@noop {} {\bibfield  {journal} {\bibinfo  {journal} {Phys. Rev. Lett.}\
  }\textbf {\bibinfo {volume} {112}},\ \bibinfo {pages} {098302} (\bibinfo
  {year} {2014})}\BibitemShut {NoStop}%
\bibitem [{\citenamefont {Seto}\ \emph {et~al.}(2013)\citenamefont {Seto},
  \citenamefont {Mari}, \citenamefont {Morris},\ and\ \citenamefont
  {Denn}}]{seto2013discontinuous}%
  \BibitemOpen
  \bibfield  {author} {\bibinfo {author} {\bibfnamefont {R.}~\bibnamefont
  {Seto}}, \bibinfo {author} {\bibfnamefont {R.}~\bibnamefont {Mari}}, \bibinfo
  {author} {\bibfnamefont {J.~F.}\ \bibnamefont {Morris}}, \ and\ \bibinfo
  {author} {\bibfnamefont {M.~M.}\ \bibnamefont {Denn}},\ }\bibfield  {title}
  {\enquote {\bibinfo {title} {Discontinuous shear thickening of frictional
  hard-sphere suspensions},}\ }\href@noop {} {\bibfield  {journal} {\bibinfo
  {journal} {Phys. Rev. Lett.}\ }\textbf {\bibinfo {volume} {111}},\ \bibinfo
  {pages} {218301} (\bibinfo {year} {2013})}\BibitemShut {NoStop}%
\bibitem [{\citenamefont {Mari}\ \emph {et~al.}(2014)\citenamefont {Mari},
  \citenamefont {Seto}, \citenamefont {Morris},\ and\ \citenamefont
  {Denn}}]{mari2014shear}%
  \BibitemOpen
  \bibfield  {author} {\bibinfo {author} {\bibfnamefont {R.}~\bibnamefont
  {Mari}}, \bibinfo {author} {\bibfnamefont {R.}~\bibnamefont {Seto}}, \bibinfo
  {author} {\bibfnamefont {J.~F.}\ \bibnamefont {Morris}}, \ and\ \bibinfo
  {author} {\bibfnamefont {M.~M.}\ \bibnamefont {Denn}},\ }\bibfield  {title}
  {\enquote {\bibinfo {title} {Shear thickening, frictionless and frictional
  rheologies in non-brownian suspensions},}\ }\href@noop {} {\bibfield
  {journal} {\bibinfo  {journal} {J. Rheol.}\ }\textbf {\bibinfo {volume}
  {58}},\ \bibinfo {pages} {1693--1724} (\bibinfo {year} {2014})}\BibitemShut
  {NoStop}%
\bibitem [{\citenamefont {Hermes}\ \emph {et~al.}(2016)\citenamefont {Hermes},
  \citenamefont {Guy}, \citenamefont {Poon}, \citenamefont {Poy}, \citenamefont
  {Cates},\ and\ \citenamefont {Wyart}}]{hermes2016unsteady}%
  \BibitemOpen
  \bibfield  {author} {\bibinfo {author} {\bibfnamefont {M.}~\bibnamefont
  {Hermes}}, \bibinfo {author} {\bibfnamefont {B.~M.}\ \bibnamefont {Guy}},
  \bibinfo {author} {\bibfnamefont {W.~C.~K.}\ \bibnamefont {Poon}}, \bibinfo
  {author} {\bibfnamefont {G.}~\bibnamefont {Poy}}, \bibinfo {author}
  {\bibfnamefont {M.~E.}\ \bibnamefont {Cates}}, \ and\ \bibinfo {author}
  {\bibfnamefont {M.}~\bibnamefont {Wyart}},\ }\bibfield  {title} {\enquote
  {\bibinfo {title} {Unsteady flow and particle migration in dense,
  non-brownian suspensions},}\ }\href@noop {} {\bibfield  {journal} {\bibinfo
  {journal} {J. Rheol.}\ }\textbf {\bibinfo {volume} {60}},\ \bibinfo {pages}
  {905--916} (\bibinfo {year} {2016})}\BibitemShut {NoStop}%
\bibitem [{\citenamefont {Singh}\ \emph {et~al.}(2018)\citenamefont {Singh},
  \citenamefont {Mari}, \citenamefont {Denn},\ and\ \citenamefont
  {Morris}}]{singh2018constitutive}%
  \BibitemOpen
  \bibfield  {author} {\bibinfo {author} {\bibfnamefont {A.}~\bibnamefont
  {Singh}}, \bibinfo {author} {\bibfnamefont {R.}~\bibnamefont {Mari}},
  \bibinfo {author} {\bibfnamefont {M.~M.}\ \bibnamefont {Denn}}, \ and\
  \bibinfo {author} {\bibfnamefont {J.~F.}\ \bibnamefont {Morris}},\ }\bibfield
   {title} {\enquote {\bibinfo {title} {A constitutive model for simple shear
  of dense frictional suspensions},}\ }\href@noop {} {\bibfield  {journal}
  {\bibinfo  {journal} {J. Rheol.}\ }\textbf {\bibinfo {volume} {62}},\
  \bibinfo {pages} {457--468} (\bibinfo {year} {2018})}\BibitemShut {NoStop}%
\bibitem [{\citenamefont {Guy}\ \emph {et~al.}(2019)\citenamefont {Guy},
  \citenamefont {Ness}, \citenamefont {Hermes}, \citenamefont {Sawiak},
  \citenamefont {Sun},\ and\ \citenamefont {Poon}}]{guy2019testing}%
  \BibitemOpen
  \bibfield  {author} {\bibinfo {author} {\bibfnamefont {B.~M.}\ \bibnamefont
  {Guy}}, \bibinfo {author} {\bibfnamefont {C.}~\bibnamefont {Ness}}, \bibinfo
  {author} {\bibfnamefont {M.}~\bibnamefont {Hermes}}, \bibinfo {author}
  {\bibfnamefont {L.~J.}\ \bibnamefont {Sawiak}}, \bibinfo {author}
  {\bibfnamefont {J.}~\bibnamefont {Sun}}, \ and\ \bibinfo {author}
  {\bibfnamefont {W.~C.~K.}\ \bibnamefont {Poon}},\ }\bibfield  {title}
  {\enquote {\bibinfo {title} {Testing the wyart-cates model for non-brownian
  shear thickening using bidisperse suspensions},}\ }\href@noop {} {\bibfield
  {journal} {\bibinfo  {journal} {arXiv preprint arXiv:1901.02066}\ } (\bibinfo
  {year} {2019})}\BibitemShut {NoStop}%
\bibitem [{\citenamefont {Gadala-Maria}\ and\ \citenamefont
  {Acrivos}(1980)}]{gadala1980shear}%
  \BibitemOpen
  \bibfield  {author} {\bibinfo {author} {\bibfnamefont {F}~\bibnamefont
  {Gadala-Maria}}\ and\ \bibinfo {author} {\bibfnamefont {A.}~\bibnamefont
  {Acrivos}},\ }\bibfield  {title} {\enquote {\bibinfo {title} {Shear-induced
  structure in a concentrated suspension of solid spheres},}\ }\href@noop {}
  {\bibfield  {journal} {\bibinfo  {journal} {J. Rheol.}\ }\textbf {\bibinfo
  {volume} {24}},\ \bibinfo {pages} {799--814} (\bibinfo {year}
  {1980})}\BibitemShut {NoStop}%
\bibitem [{\citenamefont {Lin}\ \emph {et~al.}(2015)\citenamefont {Lin},
  \citenamefont {Guy}, \citenamefont {Hermes}, \citenamefont {Ness},
  \citenamefont {Sun}, \citenamefont {Poon},\ and\ \citenamefont
  {Cohen}}]{lin2015hydrodynamic}%
  \BibitemOpen
  \bibfield  {author} {\bibinfo {author} {\bibfnamefont {N.~Y.~C.}\
  \bibnamefont {Lin}}, \bibinfo {author} {\bibfnamefont {B.~M.}\ \bibnamefont
  {Guy}}, \bibinfo {author} {\bibfnamefont {M.}~\bibnamefont {Hermes}},
  \bibinfo {author} {\bibfnamefont {C.}~\bibnamefont {Ness}}, \bibinfo {author}
  {\bibfnamefont {J.}~\bibnamefont {Sun}}, \bibinfo {author} {\bibfnamefont
  {W.~C.~K.}\ \bibnamefont {Poon}}, \ and\ \bibinfo {author} {\bibfnamefont
  {I.}~\bibnamefont {Cohen}},\ }\bibfield  {title} {\enquote {\bibinfo {title}
  {Hydrodynamic and contact contributions to continuous shear thickening in
  colloidal suspensions},}\ }\href@noop {} {\bibfield  {journal} {\bibinfo
  {journal} {Phys. Rev. Lett.}\ }\textbf {\bibinfo {volume} {115}},\ \bibinfo
  {pages} {228304} (\bibinfo {year} {2015})}\BibitemShut {NoStop}%
\bibitem [{\citenamefont {Fielding}(2007)}]{fielding2007complex}%
  \BibitemOpen
  \bibfield  {author} {\bibinfo {author} {\bibfnamefont {S.~M.}\ \bibnamefont
  {Fielding}},\ }\bibfield  {title} {\enquote {\bibinfo {title} {Complex
  dynamics of shear banded flows},}\ }\href@noop {} {\bibfield  {journal}
  {\bibinfo  {journal} {Soft Matter}\ }\textbf {\bibinfo {volume} {3}},\
  \bibinfo {pages} {1262--1279} (\bibinfo {year} {2007})}\BibitemShut {NoStop}%
\bibitem [{\citenamefont {Gillissen}\ and\ \citenamefont
  {Wilson}(2018)}]{gillissen2018modeling}%
  \BibitemOpen
  \bibfield  {author} {\bibinfo {author} {\bibfnamefont {J.~J.~J.}\
  \bibnamefont {Gillissen}}\ and\ \bibinfo {author} {\bibfnamefont {H.~J.}\
  \bibnamefont {Wilson}},\ }\bibfield  {title} {\enquote {\bibinfo {title}
  {Modeling sphere suspension microstructure and stress},}\ }\href@noop {}
  {\bibfield  {journal} {\bibinfo  {journal} {Phys. Rev. E}\ }\textbf {\bibinfo
  {volume} {98}},\ \bibinfo {pages} {033119} (\bibinfo {year}
  {2018})}\BibitemShut {NoStop}%
\bibitem [{\citenamefont {Gillissen}\ and\ \citenamefont
  {Wilson}(2019{\natexlab{a}})}]{gillissen2019effect}%
  \BibitemOpen
  \bibfield  {author} {\bibinfo {author} {\bibfnamefont {J.~J.~J.}\
  \bibnamefont {Gillissen}}\ and\ \bibinfo {author} {\bibfnamefont {H.~J.}\
  \bibnamefont {Wilson}},\ }\bibfield  {title} {\enquote {\bibinfo {title}
  {Effect of normal contact forces on the stress in shear rate invariant
  particle suspensions},}\ }\href@noop {} {\bibfield  {journal} {\bibinfo
  {journal} {Phys. Rev. Fluids}\ }\textbf {\bibinfo {volume} {4}},\ \bibinfo
  {pages} {013301} (\bibinfo {year} {2019}{\natexlab{a}})}\BibitemShut
  {NoStop}%
\bibitem [{\citenamefont {Blanc}\ \emph {et~al.}(2013)\citenamefont {Blanc},
  \citenamefont {Lemaire}, \citenamefont {Meunier},\ and\ \citenamefont
  {Peters}}]{blanc2013microstructure}%
  \BibitemOpen
  \bibfield  {author} {\bibinfo {author} {\bibfnamefont {F.}~\bibnamefont
  {Blanc}}, \bibinfo {author} {\bibfnamefont {E.}~\bibnamefont {Lemaire}},
  \bibinfo {author} {\bibfnamefont {A.}~\bibnamefont {Meunier}}, \ and\
  \bibinfo {author} {\bibfnamefont {F.}~\bibnamefont {Peters}},\ }\bibfield
  {title} {\enquote {\bibinfo {title} {Microstructure in sheared non-brownian
  concentrated suspensions},}\ }\href@noop {} {\bibfield  {journal} {\bibinfo
  {journal} {J. Rheol.}\ }\textbf {\bibinfo {volume} {57}},\ \bibinfo {pages}
  {273--292} (\bibinfo {year} {2013})}\BibitemShut {NoStop}%
\bibitem [{\citenamefont {Hinch}\ and\ \citenamefont
  {Leal}(1976)}]{hinch1976constitutive}%
  \BibitemOpen
  \bibfield  {author} {\bibinfo {author} {\bibfnamefont {E.~J.}\ \bibnamefont
  {Hinch}}\ and\ \bibinfo {author} {\bibfnamefont {L.~G.}\ \bibnamefont
  {Leal}},\ }\bibfield  {title} {\enquote {\bibinfo {title} {Constitutive
  equations in suspension mechanics. {P}art 2. {A}pproximate forms for a
  suspension of rigid particles affected by {B}rownian rotations},}\
  }\href@noop {} {\bibfield  {journal} {\bibinfo  {journal} {J. Fluid Mech.}\
  }\textbf {\bibinfo {volume} {76}},\ \bibinfo {pages} {187--208} (\bibinfo
  {year} {1976})}\BibitemShut {NoStop}%
\bibitem [{\citenamefont {Chacko}\ \emph
  {et~al.}(2018{\natexlab{a}})\citenamefont {Chacko}, \citenamefont {Mari},
  \citenamefont {Fielding},\ and\ \citenamefont {Cates}}]{chacko2018shear}%
  \BibitemOpen
  \bibfield  {author} {\bibinfo {author} {\bibfnamefont {R.~N.}\ \bibnamefont
  {Chacko}}, \bibinfo {author} {\bibfnamefont {R.}~\bibnamefont {Mari}},
  \bibinfo {author} {\bibfnamefont {S.~M.}\ \bibnamefont {Fielding}}, \ and\
  \bibinfo {author} {\bibfnamefont {M.~E.}\ \bibnamefont {Cates}},\ }\bibfield
  {title} {\enquote {\bibinfo {title} {Shear reversal in dense suspensions: The
  challenge to fabric evolution models from simulation data},}\ }\href@noop {}
  {\bibfield  {journal} {\bibinfo  {journal} {J. Fluid Mech.}\ }\textbf
  {\bibinfo {volume} {847}},\ \bibinfo {pages} {700--734} (\bibinfo {year}
  {2018}{\natexlab{a}})}\BibitemShut {NoStop}%
\bibitem [{\citenamefont {Seto}\ and\ \citenamefont
  {Giusteri}(2018)}]{seto2018normal}%
  \BibitemOpen
  \bibfield  {author} {\bibinfo {author} {\bibfnamefont {R.}~\bibnamefont
  {Seto}}\ and\ \bibinfo {author} {\bibfnamefont {G.~G.}\ \bibnamefont
  {Giusteri}},\ }\bibfield  {title} {\enquote {\bibinfo {title} {Normal stress
  differences in dense suspensions},}\ }\href@noop {} {\bibfield  {journal}
  {\bibinfo  {journal} {J. Fluid Mech.}\ }\textbf {\bibinfo {volume} {857}},\
  \bibinfo {pages} {200--215} (\bibinfo {year} {2018})}\BibitemShut {NoStop}%
\bibitem [{\citenamefont {Gillissen}\ and\ \citenamefont
  {Wilson}(2019{\natexlab{b}})}]{gillissen2019c}%
  \BibitemOpen
  \bibfield  {author} {\bibinfo {author} {\bibfnamefont {J.~J.~J.}\
  \bibnamefont {Gillissen}}\ and\ \bibinfo {author} {\bibfnamefont {H.~J.}\
  \bibnamefont {Wilson}},\ }\bibfield  {title} {\enquote {\bibinfo {title}
  {Taylor couette instability in sphere suspensions},}\ }\href@noop {}
  {\bibfield  {journal} {\bibinfo  {journal} {Phys. Rev. Fluids}\ }\textbf
  {\bibinfo {volume} {4}},\ \bibinfo {pages} {043301} (\bibinfo {year}
  {2019}{\natexlab{b}})}\BibitemShut {NoStop}%
\bibitem [{\citenamefont {Cates}\ \emph {et~al.}(1998)\citenamefont {Cates},
  \citenamefont {Wittmer}, \citenamefont {Bouchaud},\ and\ \citenamefont
  {Claudin}}]{cates1998jamming}%
  \BibitemOpen
  \bibfield  {author} {\bibinfo {author} {\bibfnamefont {M.~E.}\ \bibnamefont
  {Cates}}, \bibinfo {author} {\bibfnamefont {J.~P.}\ \bibnamefont {Wittmer}},
  \bibinfo {author} {\bibfnamefont {J.~P.}\ \bibnamefont {Bouchaud}}, \ and\
  \bibinfo {author} {\bibfnamefont {P.}~\bibnamefont {Claudin}},\ }\bibfield
  {title} {\enquote {\bibinfo {title} {Jamming, force chains, and fragile
  matter},}\ }\href@noop {} {\bibfield  {journal} {\bibinfo  {journal} {Phys.
  Rev. Lett.}\ }\textbf {\bibinfo {volume} {81}},\ \bibinfo {pages} {1841}
  (\bibinfo {year} {1998})}\BibitemShut {NoStop}%
\bibitem [{\citenamefont {Plimpton}(1995)}]{plimpton1995fast}%
  \BibitemOpen
  \bibfield  {author} {\bibinfo {author} {\bibfnamefont {S.}~\bibnamefont
  {Plimpton}},\ }\bibfield  {title} {\enquote {\bibinfo {title} {Fast parallel
  algorithms for short-range molecular dynamics},}\ }\href@noop {} {\bibfield
  {journal} {\bibinfo  {journal} {J. Comp. Phys.}\ }\textbf {\bibinfo {volume}
  {117}},\ \bibinfo {pages} {1--19} (\bibinfo {year} {1995})}\BibitemShut
  {NoStop}%
\bibitem [{\citenamefont {Cheal}\ and\ \citenamefont
  {Ness}(2018)}]{cheal2018rheology}%
  \BibitemOpen
  \bibfield  {author} {\bibinfo {author} {\bibfnamefont {O.~R.}\ \bibnamefont
  {Cheal}}\ and\ \bibinfo {author} {\bibfnamefont {C.}~\bibnamefont {Ness}},\
  }\bibfield  {title} {\enquote {\bibinfo {title} {Rheology of dense granular
  suspensions under extensional flow},}\ }\href@noop {} {\bibfield  {journal}
  {\bibinfo  {journal} {J. Rheol.}\ }\textbf {\bibinfo {volume} {62}},\
  \bibinfo {pages} {501--512} (\bibinfo {year} {2018})}\BibitemShut {NoStop}%
\bibitem [{\citenamefont {Ball}\ and\ \citenamefont
  {Melrose}(1997)}]{ball1997simulation}%
  \BibitemOpen
  \bibfield  {author} {\bibinfo {author} {\bibfnamefont {R.~C.}\ \bibnamefont
  {Ball}}\ and\ \bibinfo {author} {\bibfnamefont {J.~R.}\ \bibnamefont
  {Melrose}},\ }\bibfield  {title} {\enquote {\bibinfo {title} {A simulation
  technique for many spheres in quasi-static motion under frame-invariant pair
  drag and brownian forces},}\ }\href@noop {} {\bibfield  {journal} {\bibinfo
  {journal} {Physica A}\ }\textbf {\bibinfo {volume} {247}},\ \bibinfo {pages}
  {444--472} (\bibinfo {year} {1997})}\BibitemShut {NoStop}%
\bibitem [{\citenamefont {Chacko}\ \emph
  {et~al.}(2018{\natexlab{b}})\citenamefont {Chacko}, \citenamefont {Mari},
  \citenamefont {Cates},\ and\ \citenamefont {Fielding}}]{chacko2018dynamic}%
  \BibitemOpen
  \bibfield  {author} {\bibinfo {author} {\bibfnamefont {R.~N.}\ \bibnamefont
  {Chacko}}, \bibinfo {author} {\bibfnamefont {R.}~\bibnamefont {Mari}},
  \bibinfo {author} {\bibfnamefont {M.~E.}\ \bibnamefont {Cates}}, \ and\
  \bibinfo {author} {\bibfnamefont {S.~M}\ \bibnamefont {Fielding}},\
  }\bibfield  {title} {\enquote {\bibinfo {title} {Dynamic vorticity banding in
  discontinuously shear thickening suspensions},}\ }\href@noop {} {\bibfield
  {journal} {\bibinfo  {journal} {Phys. Rev. Lett.}\ }\textbf {\bibinfo
  {volume} {121}},\ \bibinfo {pages} {108003} (\bibinfo {year}
  {2018}{\natexlab{b}})}\BibitemShut {NoStop}%
\bibitem [{\citenamefont {Radhakrishnan}\ \emph {et~al.}(2019)\citenamefont
  {Radhakrishnan}, \citenamefont {Royer}, \citenamefont {Poon},\ and\
  \citenamefont {Sun}}]{radhakrishnan2019force}%
  \BibitemOpen
  \bibfield  {author} {\bibinfo {author} {\bibfnamefont {R.}~\bibnamefont
  {Radhakrishnan}}, \bibinfo {author} {\bibfnamefont {J.~R}\ \bibnamefont
  {Royer}}, \bibinfo {author} {\bibfnamefont {W.~C.~K.}\ \bibnamefont {Poon}},
  \ and\ \bibinfo {author} {\bibfnamefont {J.}~\bibnamefont {Sun}},\ }\bibfield
   {title} {\enquote {\bibinfo {title} {Force chains and networks: wet
  suspensions through dry granular eyes},}\ }\href@noop {} {\bibfield
  {journal} {\bibinfo  {journal} {arXiv preprint arXiv:1904.03144}\ } (\bibinfo
  {year} {2019})}\BibitemShut {NoStop}%
\bibitem [{\citenamefont {Blanc}\ \emph {et~al.}(2018)\citenamefont {Blanc},
  \citenamefont {D'Ambrosio}, \citenamefont {Lobry}, \citenamefont {Peters},\
  and\ \citenamefont {Lemaire}}]{blanc2018universal}%
  \BibitemOpen
  \bibfield  {author} {\bibinfo {author} {\bibfnamefont {F.}~\bibnamefont
  {Blanc}}, \bibinfo {author} {\bibfnamefont {E.}~\bibnamefont {D'Ambrosio}},
  \bibinfo {author} {\bibfnamefont {L.}~\bibnamefont {Lobry}}, \bibinfo
  {author} {\bibfnamefont {F.}~\bibnamefont {Peters}}, \ and\ \bibinfo {author}
  {\bibfnamefont {E.}~\bibnamefont {Lemaire}},\ }\bibfield  {title} {\enquote
  {\bibinfo {title} {Universal scaling law in frictional non-brownian
  suspensions},}\ }\href@noop {} {\bibfield  {journal} {\bibinfo  {journal}
  {Phys. Rev. Fluids}\ }\textbf {\bibinfo {volume} {3}},\ \bibinfo {pages}
  {114303} (\bibinfo {year} {2018})}\BibitemShut {NoStop}%
\bibitem [{\citenamefont {Ness}\ and\ \citenamefont {Sun}(2016)}]{ness2016two}%
  \BibitemOpen
  \bibfield  {author} {\bibinfo {author} {\bibfnamefont {C.}~\bibnamefont
  {Ness}}\ and\ \bibinfo {author} {\bibfnamefont {J.}~\bibnamefont {Sun}},\
  }\bibfield  {title} {\enquote {\bibinfo {title} {Two-scale evolution during
  shear reversal in dense suspensions},}\ }\href@noop {} {\bibfield  {journal}
  {\bibinfo  {journal} {Phys. Rev. E}\ }\textbf {\bibinfo {volume} {93}},\
  \bibinfo {pages} {012604} (\bibinfo {year} {2016})}\BibitemShut {NoStop}%
\bibitem [{\citenamefont {Denn}\ and\ \citenamefont
  {Morris}(2014)}]{denn2014rheology}%
  \BibitemOpen
  \bibfield  {author} {\bibinfo {author} {\bibfnamefont {M.~M.}\ \bibnamefont
  {Denn}}\ and\ \bibinfo {author} {\bibfnamefont {J.~F}\ \bibnamefont
  {Morris}},\ }\bibfield  {title} {\enquote {\bibinfo {title} {Rheology of
  non-brownian suspensions},}\ }\href@noop {} {\bibfield  {journal} {\bibinfo
  {journal} {Annu. Rev. Chem. Biomol.}\ }\textbf {\bibinfo {volume} {5}},\
  \bibinfo {pages} {203--228} (\bibinfo {year} {2014})}\BibitemShut {NoStop}%
\bibitem [{\citenamefont {Cwalina}\ and\ \citenamefont
  {Wagner}(2014)}]{cwalina2014material}%
  \BibitemOpen
  \bibfield  {author} {\bibinfo {author} {\bibfnamefont {C.~D.}\ \bibnamefont
  {Cwalina}}\ and\ \bibinfo {author} {\bibfnamefont {N.~J.}\ \bibnamefont
  {Wagner}},\ }\bibfield  {title} {\enquote {\bibinfo {title} {Material
  properties of the shear-thickened state in concentrated near hard-sphere
  colloidal dispersions},}\ }\href@noop {} {\bibfield  {journal} {\bibinfo
  {journal} {J. Rheol.}\ }\textbf {\bibinfo {volume} {58}},\ \bibinfo {pages}
  {949--967} (\bibinfo {year} {2014})}\BibitemShut {NoStop}%
\end{thebibliography}%
%
%
%
%
%
%
%
%
%
%
\end{document}